\documentclass[11pt]{article}
\usepackage{graphicx}
\begin{document}

\begin{center}
{\large{\bf Self-interaction Corrected Calculations of Correlated $f$-electron Systems}} \\

\vspace{0.2cm}
L. Petit$^{1,2}$, A. Svane$^{2}$, Z. Szotek$^{3}$, and W.M. Temmerman$^{3}$\\

\vspace{0.2cm}
$^{1}$ {\it Computer Science and Mathematics Division, and Center for Computational} \\
{\it Sciences, Oak Ridge National Laboratory, Oak Ridge, TN 37831, USA} \\
$^{2}$ {\it Department of Physics and Astronomy, University of Aarhus,} \\
{\it DK-8000 Aarhus C, Denmark} \\
$^{3}$ {\it Daresbury Laboratory, Daresbury, Warrington WA4 4AD, UK} \\
\end{center}

\begin{abstract}
The electronic structures of several actinide solid systems are calculated using the 
self-interaction corrected local spin density approximation. Within this scheme 
the $5f$ electron manifold is considered
to consist of both localized and delocalized states, and by varying their relative
proportions the energetically most favourable (groundstate) configuration can be
established. Specifically, we discuss elemental Pu in its $\delta$-phase,
PuO$_2$ and the effects of addition of oxygen, the series of actinide monopnictides and 
monochalcogenides, and the UX$_3$, X= Rh, Pd, Pt, Au, intermetallic series.
\end{abstract}

\section{Introduction}

The quantum-mechanical understanding of the physics of actinide compounds
presents a challenge due to the intricate nature of the 
partially filled $5f$-shell. Compared to the rare-earths, for which the $4f$-states 
are most often completely localized, e.g. exhibiting atomic-like multiplet structure, 
the $5f$ states in the actinides are less inert and can play a significant 
role in bonding, depending on the specific actinide element and the chemical 
environment. This  is most convincingly demonstrated in the elemental metals, for 
which a localization transition occurs when going from Pu to Am. In the early 
actinides, Th, Pa, U, Np, and the $\alpha$-phase of 
Pu, the relatively delocalized 5$f$-electrons 
actively contribute to bonding, and their atomic volumes decrease in a parabolic
fashion, similarly to the behavior seen across the transition metal series.\cite{johansson}
In Am, the $f$-electron localization is accompanied by an abrupt $\sim 50 \%$ increase in 
the atomic volume, and for the heavier elements, Cm, Bk and Cf, the specific
volume either 
remains constant or decreases only slightly. Pu lies at the borderline, and its very 
complex phase diagram implies that the $f$-electron properties are of particularly
intricate nature. Depending on the chemical properties of the ligands, the
actinide compounds may exhibit different degrees of $f$-electron localization for 
the same actinide element. 

Over the past 30 years, the local spin density (LSD) and semi-local (generalized 
gradient - GGA) approximations to density functional theory\cite{rojones,perdew-chevary} 
have proven very useful and accurate in describing bonding properties of solids 
with weakly correlated electrons, demonstrating that the cohesive energy data 
for the homogeneous 
electron gas, that underlie these approximations, are representative of 
the conduction states in real materials. However, when  4$f$-electrons are involved, 
the atomic picture with localized partially filled $f$-shells is usually a better starting 
point for calculations. The most well known extensions of LSD, capable of describing 
electron localization, 
include the self-interaction corrected (SIC)-LSD,\cite{temmerman-svane} 
LDA+U,\cite{anisimov} and orbital polarization methods.\cite{uppsala}

It is possible to get a reasonable description of rare-earth materials with the
LSD method by including a partially occupied $f$-shell into the core and projecting
out the $f$-degrees of freedom from the valence bands.\cite{delin} In such
calculations, a combination of density functional theory with input from experimental 
data is used to describe bonding electrons and the atomic $f^n$ configuration, 
respectively. The SIC-LSD method can be viewed as effectively including an integer 
number of $f$-electrons in the core, however without restricting the unoccupied 
$f$-degrees of freedom. The localized $4f$ electrons in the rare earth metals and 
compounds have been well described 
by the SIC-LSD method.\cite{strange,temmerman-szotek,svane-temmerman} 
Intermediate valent Yb 
compounds have been described\cite{svane-temmerman} 
as a localized $f^{13}$ configuration plus a narrow $f$-band state pinned to the 
Fermi level. The free Yb atom is divalent with a completely filled $f^{14}$ shell.
Thus, the 
destabilization of the localized $f$-manifold, which occurs in the solid state,
is described in the SIC-LSD method by introducing two kinds of 
$f$-electrons.\cite{gschneider} An integer number of $f$-electrons are localized while    
  a non-integer 
number of hybridized band-$f$ electrons is  determined by the self-consistent position of 
the Fermi level. 
A similar picture has emerged from calculations applying the
LDA+U approach to Tm compounds.\cite{antonov} 
The number of localized $f$-electrons leads to a definition of valency 
of the actinide ions, given as the integer number of electrons available for band 
formation. Due to a substantial $f$-character of the valence bands, this valency is not the same 
as that determined by the total $f$-electron count, which includes both localized
and itinerant $f$-electrons, and which is usually non-integral. Therefore, the most stable 
actinide valency in a given compound is determined by the balance between the localization
energy and band formation energy (hybridization energy).

In section 2 of the present paper, the SIC-LSD method is briefly described. 
In section 3, the results 
for selected cases are presented and discussed,
notably the $\delta$-phase of Pu, the actinide monopnictides and monochalcogenides,
the PuO$_2$ compound, and the UX$_3$ intermetallics.
The paper is concluded in section 4.

\section{The SIC-LSD scheme}

\begin{figure}[t]
\begin{center}
\includegraphics[scale=0.39,angle=-90]{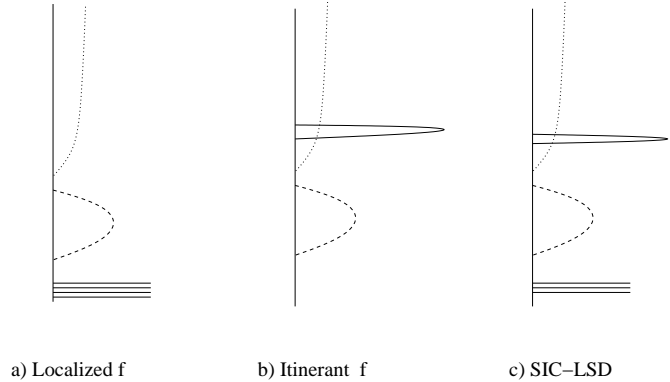}\\
\caption{Schematic representation of the density of states in the SIC-LSD
approach for an AcX compound.
a) LSD calculation with all $f$
electrons treated as inert core electrons, b) LSD calculation with all $f$
electrons treated as band states, and c) SIC-LSD calculation with both
localized and delocalized $f$ states. The dashed line represents the ligand $%
p$-band, while the broad actinide $d$-band is given by the dotted line, and $%
f$-states are shown with full line.
\label{SchemaDos}
 }
\end{center}
\end{figure}

The electronic configuration of the actinide atoms is 
[Rn]$5f^36d^17s^2$ for U, \newline [Rn]$5f^46d^17s^2$ for Np,
[Rn]$5f^67s^2$ for Pu, [Rn]$5f^77s^2$ for Am, and [Rn]$5f^76d^17s^2$ for Cm.
In the solid state, the relative proportions of $s$, $d$ and $f$ electrons
will change due to hybridization and charge transfer. The $f$ orbitals participate 
in bonding through their overlap with the $f$ and $d$ orbitals on neighbouring 
actinide ions, as well as, with the valence orbitals of the ligands.
Either of these interactions lead to a broadening of the atomic $f$-level
into an $f$ resonance, which one might hope to describe in two opposite
limits, either 
as a hybridized band (as in the standard LSD picture), or as
an atomic-like delta-function (by including a partially filled
$f^n$ shell into the atomic core and decoupling all the other $f$-degrees of 
freedom, i.e., completely ignoring a possible $f$-electron contribution to bonding). 
These two extremal scenarios 
are depicted schematically in Figs. \ref{SchemaDos}a and \ref{SchemaDos}b, 
respectively, while the SIC-LSD scenario, which can be viewed as an interpolation 
between the two, is displayed in Fig. \ref{SchemaDos}c.

In the SIC-LSD approach \cite{perdew} the LSD total energy functional is 
corrected for the spurious self-interaction of each occupied state $\alpha$:
\begin{equation}
E^{SIC}=E^{LSD}-\sum_{\alpha}^{occ.} \delta_{\alpha}^{SIC},
\end{equation}
where the self-interaction correction, $\delta_{\alpha}^{SIC}$, for a given state 
$\alpha$, is defined as the sum of the Hartree and exchange-correlation energies:
\begin{equation}
\delta_{\alpha}^{SIC}=U[n_{\alpha}]+E_{xc}^{LSD}[n_{\alpha}] .
\end{equation}
This correction vanishes for an itinerant state, and therefore the SIC-LSD
functional for such a state coincides with the conventional LSD functional. 
To benefit from the self-interaction correction, an electron state needs to 
spatially localize, which costs band formation energy due to loss of hybridization. 
Whether this is favorable depends on the relative values of the hybridization energy 
and the self-interaction correction energy. Hence, the latter is identified with 
the localization 
energy. The rationale behind the functional in Eq. (1) is that for a delocalized 
electron the interaction with a given atom is well described by the mean-field LSD
potential. In contrast, the appropriate potential for a localized electron, 
due to a large Wigner delay time,
will be corrected for the fact that other electrons on that atom 
rearrange in response to the presence of this localized electron.
The self-interaction correction depends on the spatial distribution of the $f$
orbital, while the hybridization energy depends on the overlap of a given
$f$ orbital with the $f$ and $d$ orbitals on the neighbouring actinide sites, 
and valence orbitals on the ligand sites. The $f$-electron which has become localized, by 
benefiting from the self-interaction correction, can no longer hybridize with the
conduction electron bands to give rise to any band-related features or valency fluctuations. 
However, the $f$-states which have not been explicitly localized can hybridize with 
the conduction electrons and form fully or partially occupied bands 
(see Fig. \ref{SchemaDos}c). 

In the SIC-LSD formulation one deals with two types of $f$ electrons, the
localized and hybridized $f$ electrons, as first implicated by Gschneider\cite{gschneider}
in relation to rare earths. By assuming different $f^{n}$ configurations of localized
electrons, various valency configurations can be realized and studied in detail. 
Within SIC-LSD the valency is defined as
the integer number of actinide valence electrons which are available for band formation, i. e.:
\begin{equation}
N_{val}=Z -N_{core}-N_{SIC}.
\end{equation} 
Here $Z$ is the atomic number, $N_{core}$ is the number of core (and semi-core) 
electrons (which for actinides is 86), and $N_{SIC}$ is the number of localized $f$-electrons 
on the actinide sites.  
Thus, e.g. a trivalent configuration of the actinide ions 
U$^{3+}$, Np$^{3+}$, Pu$^{3+}$, Am$^{3+}$,
and Cm$^{3+}$, is realized by localizing three, ($f^3$ configuration),
four ($f^{4}$), five ($f^{5}$), 
six ($f^{6}$), and seven ($f^{7}$) $f$ 
electrons on the respective actinide atoms. For a given $f^n$ configuration, the 
minimum in the total energy as a function of lattice parameter determines the equilibrium 
lattice constant. By comparing the total energy minima for different $f^{n}$ configurations,
the global groundstate configuration and lattice constant can be determined. 
In selecting the $f^n$ configuration the Hund's rules are usually
followed by alligning spins
and maximizing the orbital moment in the direction opposite to the spins
(for less than half-filled shells, or in parallel to the spins for more than half-filled shells).  
During the
iterations towards self-consistency the localized states are allowed to relax, although
generally they do not change much.

The SIC-LSD scheme has been implemented\cite{temmerman-svane} within the tight-binding 
linear-muffin-tin orbitals (TB-LMTO) method.\cite{oka} The actinide semi-core $6s$ and 
$6p$ states have been described with a separate energy panel. 
Spin-orbit coupling has been fully included in the self-consistency cycles. For 
simplicity, for systems discussed here, we have assumed ferromagnetic arrangement of the magnetic moments.

\section{Results and Discussion}

\subsection{$\delta$-Plutonium}

Electronic structure calculations treating $f$-electrons as band states describe 
quite succesfully the equilibrium volumes of the early actinide metals.\cite{jones} 
In an early study of Am, Skriver et al.\cite{skriver} found the $f$-electron localization,
signalled by the onset of spin-polarization, 
giving rise to an almost full, and hence nonbonding, 
spin polarized $f^7$ band. Also, 
the high pressure phases of Am have been succesfully described 
by the standard LSD theory.~\cite{oeriksson,psoederlind} 
Recently, the SIC-LSD method was
applied to the series of actinide metals\cite{petit} from Np to Fm, correctly describing 
the itinerant nature of Np, the trivalency of Am, Cm, Bk and Cf, and the shift to
divalency in Es and Fm. Pu turns out to be the most delicate case, 
being situated on the 
borderline between the itinerant and well localized actinides. The groundstate
$\alpha$-phase is well reproduced by LDA calculations\cite{jones}, but the high
temperature $\delta$-phase is peculiar. The crystal structure is high-symmetry fcc, 
it has the largest specific volume of all Pu allotropes (25 \% larger than that of
$\alpha$-Pu), and the thermal expansion coefficient is negative. 
It has long been recognized that these facts are connected to a higher degree of
localization of the $f$-electrons in the $\delta$-phase, but on the other hand the 
volume is still $\sim 16 \%$ smaller than that of Am.

\begin{figure}
\begin{center}
\includegraphics[scale=.50]{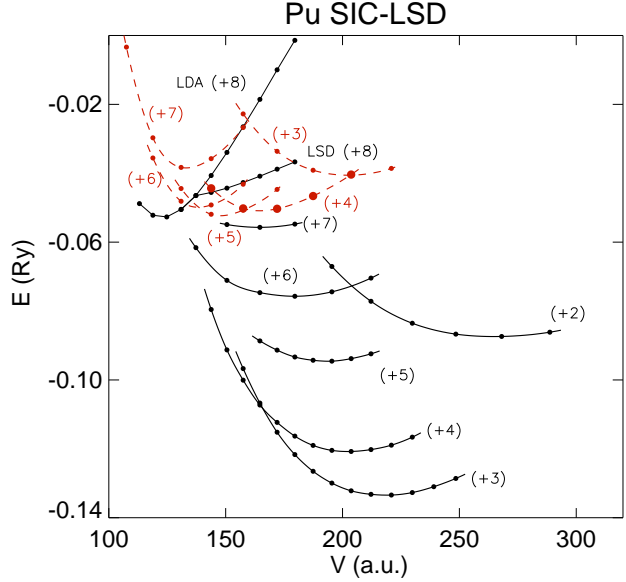}\\
\caption{Total energy for $\delta$-Pu. 
Dashed curves assume $j-j$ coupled localized $f^n$ shell,
black curves assume $L-S$ coupling.
\label{Pu-fig}   
}
\end{center}
\end{figure}

We have investigated the $\delta$-phase of Pu within the SIC-LSD approach.
The total energy of a number of localization scenarios are summarized in Figure \ref{Pu-fig}.
In constructing the localized $f^n$ shells we have considered either $L-S$ coupling
(black curves) or $j-j$ coupling (dashed curves) of the $f$-states. 
In the former case, all $f$-states are taken to be
eigenstates of $l_z$ and $s_z$, with $s_z$ quantum number $1/2$ corresponding to spin-up, 
in accordance with Hund's first rule, and $l_z$ quantum numbers occupied in the sequence
$-3,-2,-1,0,1,2,3$ to comply with Hund's second rule (for less than half filled shells).
In the latter case, one-electron $f$-states are taken as eigenstates of $j^2$ and $j_z$,
with $j=5/2$ and $j_z$ quantum number occupied in the sequence
$-5/2,5/2,-3/2,3/2,-1/2,1/2$. It is important to stress that these are only the
starting configurations of the localized states. Ultimately, the localized states
are determined  self-consistently by the SIC-LSD total energy minimization, but in
practice the symmetry of the initial state tends to be preserved during iterations
towards self-consistency. In other words there are energy barriers for a SIC state to
drastically alter its symmetry.

One notices a significant energy gain for the configurations with large spin localized
$f^n$ shells. The lowest energy is found for a localized $L-S$ coupled $f^5$ shell,
corresponding to trivalent Pu atoms.\cite{petit} The equilibrium volume is 218 $a_0^3$,
which is 30 \% larger than the experimental volume of $\delta$-Pu. Clearly, this is not
the appropriate representative of $\delta$-Pu. One more localized $f$-electron leads to
an even larger volume and also a larger total energy, while fewer localized electrons
do lead to smaller equilibrium volume but also larger total energy. The best agreement
with the experimental volume within the $L-S$ coupling scheme is obtained for the
$f^2$ localized scenario. The $j-j$ coupling scheme leads to a
completely different picture. In this case the scenarios with localized $f^2$, $f^3$,
and $f^4$ shells are almost degenerate in energy, with $f^3$ having the lowest energy,
and an equilibrium volume $12 \%$ {\it smaller} than the experimental $\delta$-Pu volume.
The conclusion to be drawn here is that the LSD provides a poor account of the energetics of
$\delta$-Pu: the exchange energy gained by the formation of large aligned spins 
is overestimated and leads to the wrong representation of the ground state. By taking $j-j$
coupled localized shells, one artificially turns the spin-density contribution to the
total energy off, and an improved  description is obtained. This does {\bf not} mean that
the $j-j$ coupled Pu $f^3$ shell is the correct ground state of Pu. Rather, the study
demonstrates  that more complicated ground states are called for. Within the restricted
one-electron picture the $j-j$ coupled localized shell is a better representation of
the true ground state. Firstly, the true
ground state must describe appropriately the spin fluctuations leading to the quenching
of the Pu moment, secondly it is also likely that fluctuations in the number of
localized $f$-electrons are needed. It is important to stress that $\delta$-Pu is a
special case in the actinide series. When going to Am, the $L-S$ ground state obtained
with the SIC-LSD approach is quite adequate, leading to a localized $f^6$, $M_S=3$, $M_L=3$,
i.e. $J=0$ ground state. The equilibrium volume is $\sim 8 \%$ larger than the
experimental volume, which is acceptable. The $j-j$ coupled ground state is also 
$f^6$, $J=0$, in this case with a volume 
only $3\%$ larger than the experimental volume, but there
is  not such a drastic difference between the two representations for the well localized Am
case as for Pu.
Hence, the failure of the SIC-LSD in describing the highly correlated $\delta$-phase
of Pu has been traced back to the large magnetic 
moment on Pu, persisting in the SIC-LSD description. By artificially setting the exchange 
interaction to zero, a much improved lattice constant has been obtained,
as also found by Refs. \cite{savrasov} and \cite{eriksson-pu}. 
Since experiments find Pu to be non-magnetic, one must conclude that the mean-field 
approaches of LSD and SIC-LSD overestimate the tendency towards magnetic moment
formation, by not taking into account quantum fluctuations in the $f$-shell. Recently,
Savrasov {\it et al.}\cite{kotliar} have presented a  promising way of treating
dynamical fluctuations and applied it successfully to $\delta$-Pu.

\subsection{Actinide Monopnictides and Monochalcogenides}

In the actinide monopnictides and monochalcogenides, 
which all crystallize 
in the NaCl structure at ambient conditions, the actinide-actinide
separations are larger than in the elemental metals, and the tendency towards 
$f$-electron localization can already be observed from Np compounds
onwards.~\cite{hill,santini,huray,vog-mat,vog-mat2,kalvius,kalvius2}


Here,         we present the SIC-LSD electronic structures calculations of 
the monopnictides and monochalcogenides of U, Np, Pu,\cite{petit2}
Am,\cite{petit1} and Cm. 
Figure 3 displays the calculated actinide ground state configurations through the series.

\begin{figure}
\begin{center}
\includegraphics[scale=0.35]{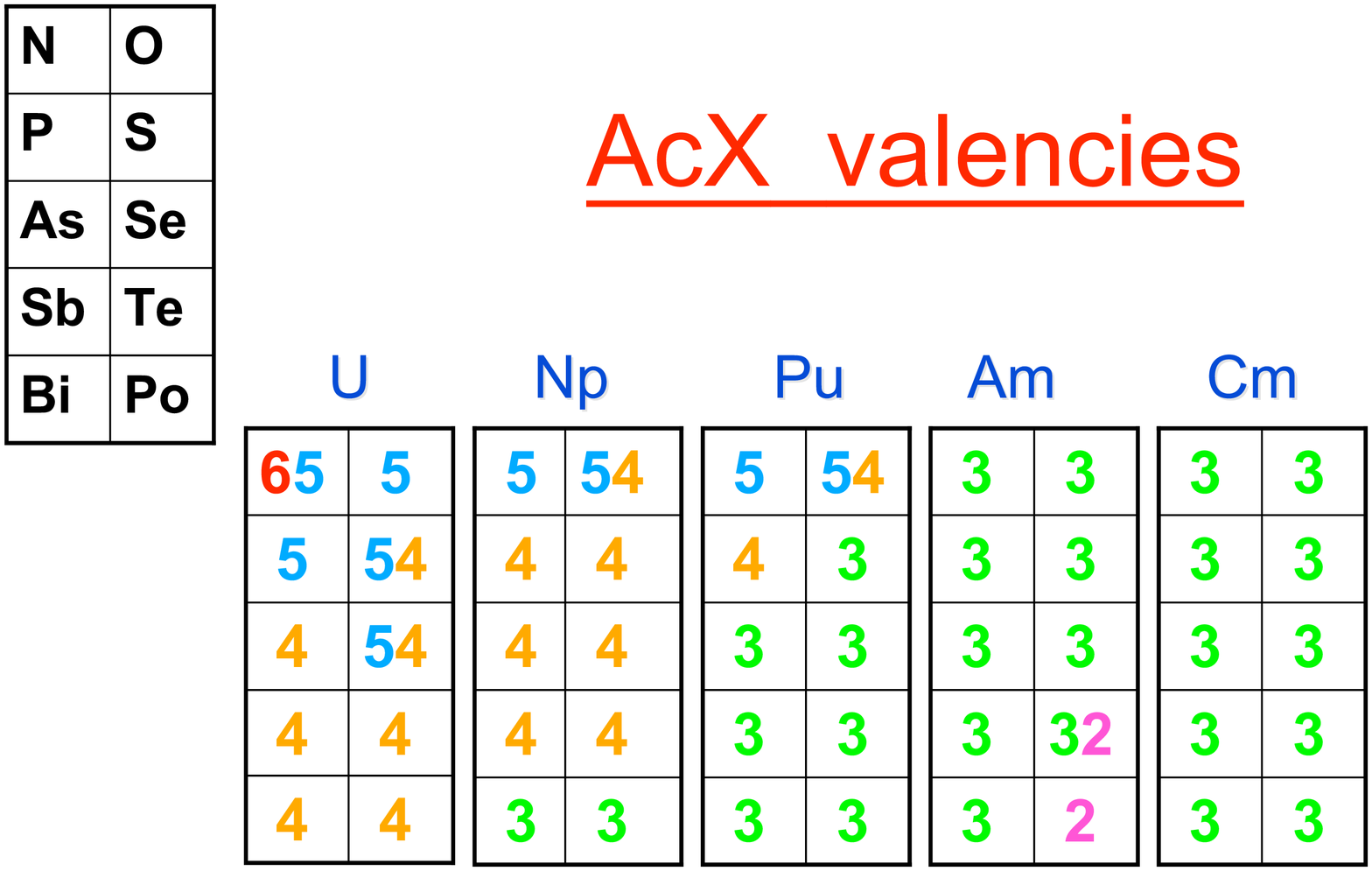}\\
\caption{ Trends in localization through the AcX series. 
For each actinide, Ac= U, Np, Pu, Am, Cm,
 a block of 10 ligands are considered: the pnictides 
X=N, P, As, Sb,    Bi, and the
chalcogenides X=O, S, Se, Te and Po. The numbers designate the calculated Ac 
valency (according to Eq. (3)) for that particular AcX compound. Where two numbers
are given, the corresponding valencies are degenerate.
\label{AcX}
}
\end{center}

\end{figure}

The calculations reveal clear trends towards more and more actively bonding
$f$-electrons for a) lighter actinides, and b) lighter ligands. For the 
lighter actinides, the $f$-orbitals are more extended leading to larger overlaps
with their nearest neighbours and smaller self-interaction corrections, both of these
effects are favoring band formation. For the lighter ligands, in particular N and O, both
the volume is decreased and ionicity is larger, the first of these effects leading to
larger direct actinide-actinide overlap, and the latter effect favoring charge
transfer. 

The Cm compounds are the most localized systems, all exhibiting Cm in the trivalent
$f^7$ configuration. The $f^7$ shell is so stable that 
variations of the ligand cannot disrupt its stability and scenarios with either
one more or less localized $f$-electron have distinctly higher energies. 
Trivalency prevails in the Am compounds, but the stability of the $f^7$ shell
causes the divalent Am state to be important in AmTe and AmPo. In the Pu compounds,
the trivalent state also dominates, but for the lighter ligands $f$-electron 
delocalization sets in. In the Np compounds the tetravalent state dominates, 
while in the U compounds pentavalent states occur for the lighter ligands.

\begin{figure*}[t]
\begin{tabular}{cc}
\includegraphics[width=70mm]{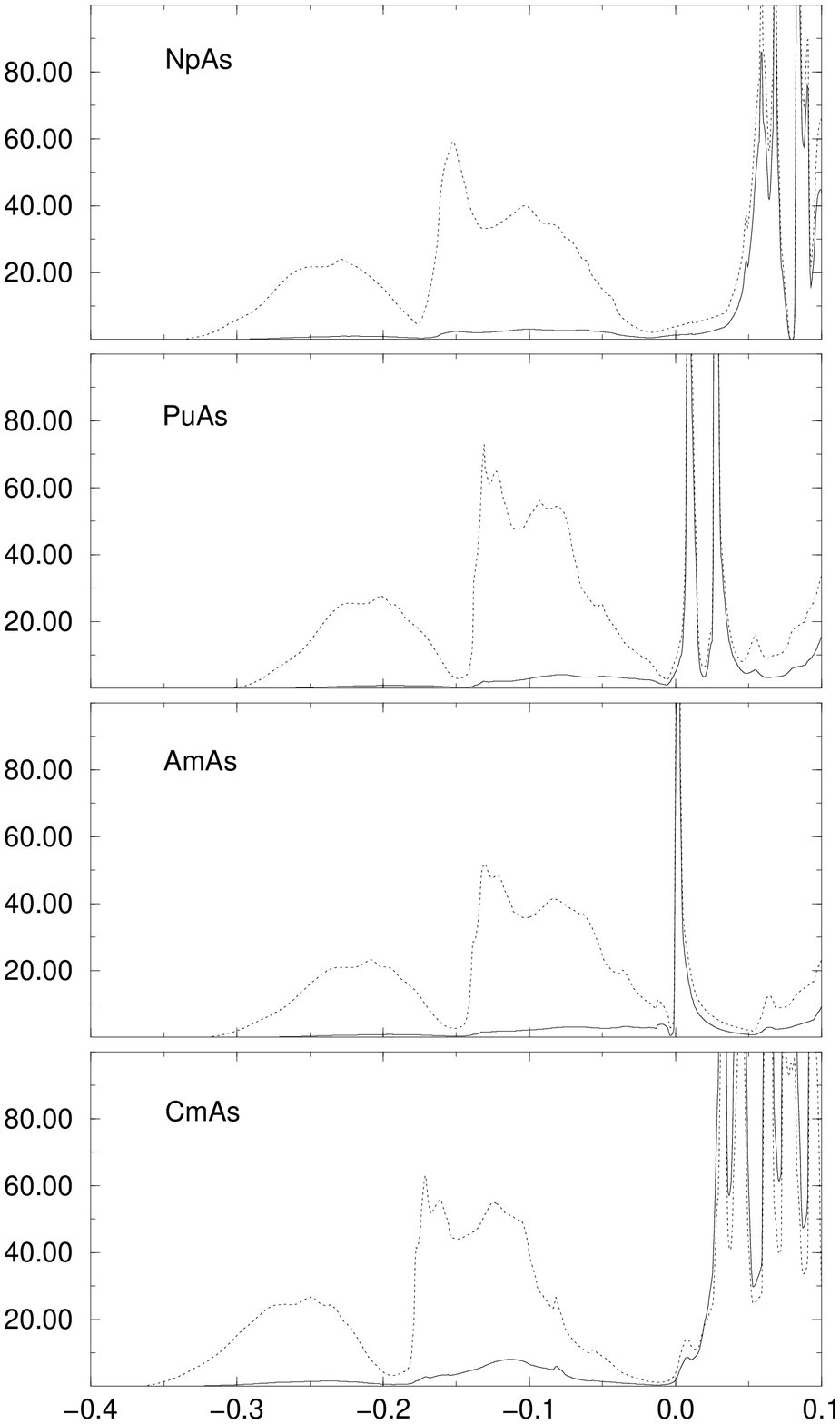}&
\includegraphics[width=70mm]{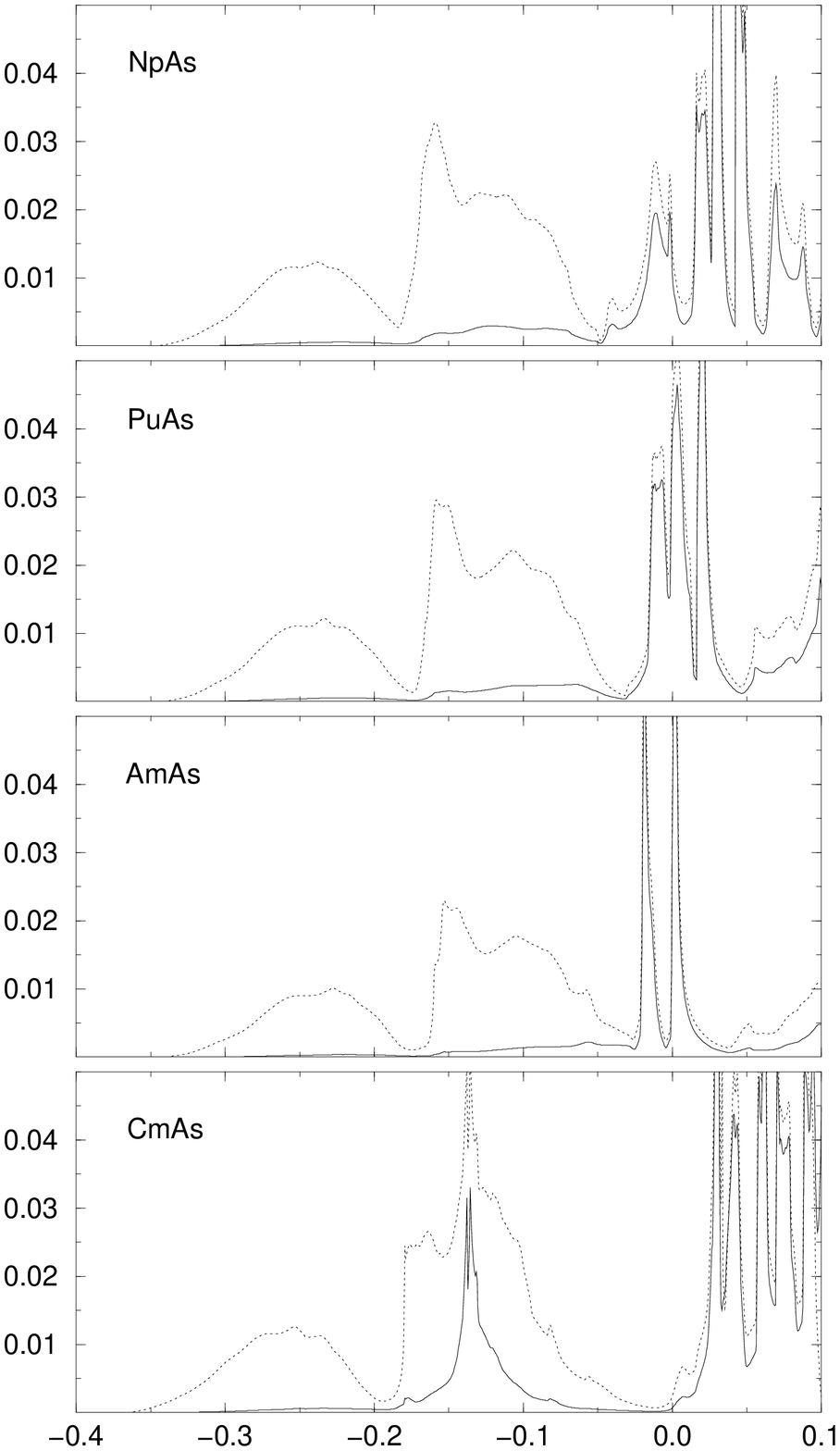} 
\end{tabular}

\caption{ Densities of states (in states/(Ry formula unit)) 
for the actinide arsenides NpAs, PuAs, AmAs, CmAs, 
with  a trivalent (left), or tetravalent (right) actinide ion, 
respectively. The solid and dotted lines represent the $f$ projected- and  total 
densities of states, respectively. The energies are given in Ry, with the Fermi level
at energy zero.
\label{DOS}
}

\end{figure*}

The densities of states of the actinide arsenides are shown in Fig. \ref{DOS},
with both trivalent and tetravalent actinide ions. In the trivalent case, the non-localized
$f$-degrees of freedom give rise to narrow unoccupied bands above the Fermi level. In the
tetravalent case the additional delocalized $f$-electron appears as an extra $f$-band. In Cm,
this band appears far below the Fermi level, while in Am, Pu and Np this band lies just below
the Fermi level. The band formation energy due to this extra band is sufficiently large in NpAs
to outweigh the localization energy, and the tetravalent configuration becomes the ground state.

%

\begin{figure}
\begin{center}
\includegraphics[scale=0.50,angle=-90]{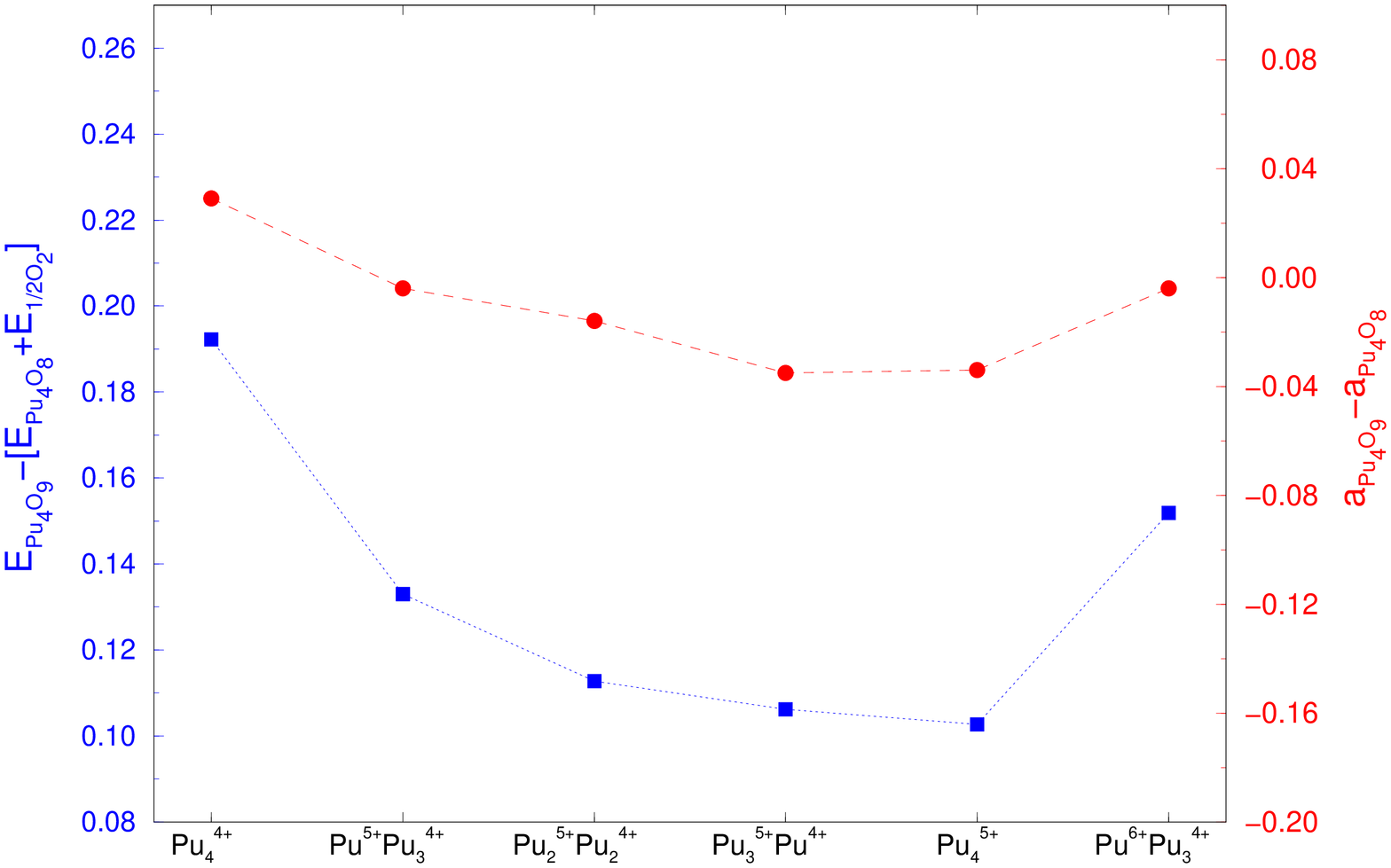}\\
\caption{Total energy minima (squares and left-hand side axis)
and corresponding theoretical equilibrium lattice constants (circles and right-hand side
axis)
for different Pu $f-$configurations of Pu$_4$O$_9$ supercell. The energy (in Ry units)
and lattice parameter data (in \AA\ units)
are given with respect to the corresponding
values of Pu$_4$O$_8$+$\frac{1}{2}$O$_2$.
\label{PuO2-fig}
}
\end{center}
\end{figure}

\subsection{Plutonium Dioxide}

PuO$_2$
is the most favored compound for storage of Pu from nuclear waste. In the stoichiometric compound Pu
is tetravalent with a localized $f^4$ shell, filled O $p$ bands and a large insulating gap. Recently,
the chemical inertness of 
PuO$_2$ has been questioned, in particular whether reactions with water could lead to further 
oxidation and the formation of 
PuO$_{2+x}$.\cite{haschke}
We investigated\cite{petit-science} the 
PuO$_{2+x}$ system  with the SIC-LSD approach by constructing a supercell with four 
PuO$_2$ units together with an additional O atom in the interstitial region, thus forming a model
of a 
PuO$_{2.25}$
compound. The interesting conclusion to be drawn from the total energy as a function of $f$-localization
(fig. \ref{PuO2-fig}) is that the nearest neighbour Pu atoms of the interstitial O transform to the 
pentavalent configuration by delocalizing one $f$-electron, which is donated to the extra O to 
form hybridized states, occuring in the gap-region of the pure 
PuO$_2$ compound.
Similarly, an O vacancy in 
PuO$_2$
will lead to the formation of trivalent $f^5$ Pu ions in the vicinity of the vacancy. In effect, 
the localized $f^n$ shell of Pu
acts as a reservoir for absorbing or releasing electrons to be accomodated by
the chemical bonds of the O atoms.
The lattice constant of the 
PuO$_{2+x}$ system almost does not vary with $x$ due to two opposing effects (Fig. \ref{PuO2-fig}). The
added O per se leads to lattice expansion, but the additional bonding due to the formation
of pentavalent Pu causes the lattice to contract.

\subsection{UX$_3$ compounds, X=Rh, Pd, Pt, Au}

The sequence of Uranium intermetallics, UX$_3$, X=Rh, Pd, Pt and Au are interesting due to their
variation in metallic properties.
This is well reproduced in the SIC-LSD approach.\cite{petit-prl}
The U configuration changes from $f^0$ in URh$_3$ to
$f^1$ in UPt$_3$ and to $f^2$ in UPd$_3$ and UAu$_3$. 
This is due to hybridization of the ligand $d$-band, which in
URh$_3$ is not completely occupied, with the U $f$-electrons to form hybridized bands rather
than non-bonding localized states. 
In UPd$_3$ and UAu$_3$ the $d$-band is full, and the U $f$-electrons can not contribute further to the bonding,
while UPt$_3$ is the borderline case, where the $d$-band is full but sufficiently close to the Fermi level
that the $f$-electrons can hybridize in, i.e., the U $f$-manifold is situated in between the fully delocalized and
fully localized scenarios, in good accord with the observation of heavy fermion properties of this compound.

\section{Conclusions}

In summary, we have reviewed the electronic properties of a number of $f$-electron systems, as obtained
within SIC-LSD approach. We have demonstrated that this approach is well suited to describe trends
regarding lattice parameters and valencies of these systems. It works especially well for systems
with well localized $f$-shells. It also indicates that the ground state of $\delta$-Plutonium
is more complex than the SIC-LSD can describe
and thus underlines the need of developing a dynamic generalization of the SIC-LSD approach.

\section{Acknowledgments}
This work has been
partially funded by the Training and Mobility Network on `Electronic Structure 
Calculation of Materials Properties and Processes for Industry and Basic Sciences'
(contract:FMRX-CT98-0178) and by the Research Training Network on 
`Ab-initio Computation of Electronic Properties of f-electron Materials'.

\end{document}